\documentclass[12pt]{article}
\usepackage{amssymb}
\usepackage{amsmath}
\def\d{\delta}

\def\la{\lambda}

\def\be{\begin{equation}}
\def\ee{\end{equation}}
\def\arr{\begin{array}{rll}}
\def\ea{\end{array}}
\def\bea{\begin{eqnarray}}
\def\eea{\end{eqnarray}}

\def\N2{$N{=}2$}

\def\>{\rangle}
\def\<{\langle}
\def\+{\dagger}
\def\={\ =\ }

\begin{document}
\renewcommand{\thefootnote}{\arabic{footnote}}
\begin{titlepage}
\setcounter{page}{0}
\begin{flushright}
$\qquad$ \\
\end{flushright}
\vskip 1.4cm
\begin{center}
{\LARGE\bf New realizations of $\,\mathcal{N}=2$ $l$-conformal}\\
\vskip 0.8cm
{\LARGE\bf Newton-Hooke superalgebra}\\
\vskip 2cm
$
\textrm{\Large Ivan Masterov\ }
$
\vskip 1cm
{\it
Laboratory of Mathematical Physics, Tomsk Polytechnic University, \\
634050 Tomsk, Lenin Ave. 30, Russian Federation}
\vskip 0.7cm
{E-mail: masterov@tpu.ru}

\end{center}
\vskip 1.2cm
\begin{abstract} \noindent
By applying Niederer--like transformation,
we construct a representation of the $\,\mathcal{N}=2$ $l$-conformal Newton-Hooke superalgebra for the case of a negative cosmological constant in terms of linear differential operators as well as its dynamical realization.  Another variant of $\,\mathcal{N}=2$ supersymmetric Pais-Uhlenbeck oscillator for a particular choice of its frequencies is proposed. The advantages of such realizations as compared to their analogues introduced in [J. Math. Phys. 53 (2012) 072904], [J. Math. Phys. 56 (2015) 022902] are discussed.
\end{abstract}


\vskip 0.7cm


\end{titlepage}

\noindent
{\bf 1. Introduction}
\vskip 0.5cm

The Newton-Hooke (NH) algebra can be obtained from (anti) de Sitter algebra by nonrelativistic contraction procedure \cite{Bacry}. The interest in the NH algebra was originally motivated by its relations to physical systems  in NH spacetime, i.e. in spacetime with nonrelativistic cosmological attraction or repulsion \cite{Bacry}-\cite{Dubois_2}.

In 1973, Niederer showed that the Schr\"{o}dinger equation for the harmonic oscillator is invariant under transformations which form the NH group enlarged by dilatations and special conformal transformations \cite{Niederer}. Moreover, it was demonstrated that conformally extended NH group is locally isomorphic to the symmetry group of the free Schr\"{o}dinger equation \cite{Niederer_1}. It was then shown that the symmetries of the equations mentioned above are related by a change of coordinates which affects the time. In modern literature this coordinate transformation and its analogues are referred to as Niederer's transformation \cite{Horvathy}-\cite{Masterov_3}.

The systematic study of nonrelativistic conformal algebras \cite{Negro_1,Negro_2} shows that the NH algebra can be conformally extended in different ways. In general, the family of such extensions is called the $l$-conformal NH algebra, where $l$ is a positive integer or a half-integer parameter. Structure relations of this algebra involve the nonrelativistic cosmological constant $\Lambda=\pm\frac{1}{R^2}$, where $R$ is the characteristic time which is proportional to the radius of the parent (anti) de Sitter spacetime \cite{Gibbons}. When $\Lambda$ tends to zero, one reproduces the so-called $l$-conformal Galilei algebra \cite{Negro_1,Negro_2,Henkel} which, however, is isomorphic to its NH counterpart. Indeed, the linear change of the basis in the $l$-conformal Galilei algebra
\bea\label{change_2}
K_{-1}\rightarrow K_{-1}\pm\frac{1}{R^2}K_{1},
\eea
where $K_{-1}$ is the generator of time translations, $K_{1}$ is the generator of special conformal transformations, leads to the structure relations of the $l$-conformal NH algebra with negative (upper sign) or positive (lower sign) cosmological constant \cite{Galajinsky_3}. However, if one is interested in dynamical realizations, the change (\ref{change_2}) alters the Hamiltonian and, consequently, the dynamics of a systems (see related discussions in Refs. \cite{Horvathy}, \cite{Galajinsky_3}).

The extensive investigation of the nonrelativistic AdS/CFT correspondence\footnote{There is extensive literature on this subject. For a review and further references see e.g. Ref. \cite{Dobrev}.} stimulates the interest in the $l$-conformal NH algebra \cite{Galajinsky_3}, \cite{Masterov_1}, \cite{PU}-\cite{Masterov_3}, \cite{Liu}-\cite{Andrzejewski_1}. In particular, in recent works \cite{PU,Galajinsky_1} (see also Refs. \cite{Andrzejewski}, \cite{Andrzejewski_1}) it was demonstrated that the $l$-conformal NH symmetry corresponding to negative cosmological constant underlies the Pais-Uhlenbeck oscillator\footnote{The invariance of the Pais-Uhlenbeck oscillator under the $l$-conformal NH algebra was anticipated in Ref. \cite{Gomis}.} \cite{Pais} for a particular choice of its frequencies.

The Pais-Uhlenbeck oscillator is arguably the most popular higher derivative mechanical system. In general, quantum description of higher derivative systems faces problems. One either reveals ghosts or the unbounded from below spectrum, or
the absence of the ground state \cite{Pais}. However, one can try to avoid the troubles by switching over to a supersymmetric extension \cite{Smilga_2,Smilga_1}. Indeed, suppose that the quantum system with the hermitian Hamiltonian $K_{-1}$ possesses the supercharges $G_{-\frac{1}{2}}$, $\bar{G}_{-\frac{1}{2}}$ which are hermitian conjugates of each other. Then all eigenvalues of the Hamiltonian are non-negative and the ground state exists provided the (anti)commutation relations in the superalgebra hold the conventional form (see a related discussion in Ref. \cite{Smilga_1})
\bea
\{G_{-\frac{1}{2}},G_{-\frac{1}{2}}\}=0,\qquad \{G_{-\frac{1}{2}},\bar{G}_{-\frac{1}{2}}\}=2 K_{-1},\quad \{\bar{G}_{-\frac{1}{2}},\bar{G}_{-\frac{1}{2}}\}=0.
\nonumber
\eea

These considerations motivate the investigation of the $\,\mathcal{N}=2$ supersymmetric extensions of the $l$-conformal NH algebra which has been initiated in our recent works \cite{Masterov_1}, \cite{Masterov_3}.
The purpose of this paper is to modify Niederer's transformations in \cite{Masterov_1}, \cite{Masterov_3} so as to derive new representations of the $\mathcal{N}=2$ $l$-conformal Galilei superalgebra in NH spacetime with negative cosmological constant. In particular, we construct new representation in terms of linear differential operators in Section 3. A new dynamical realization is obtained in Section 4. We summarize our results and discuss further possible developments in the concluding Section 5. Symmetry transformations of the action functionals exposed in Section 4 are gathered in Appendix.

\vskip 0.5 cm
\noindent

\noindent
{\bf 2. The NH bases of the $\,\mathcal{N}=2$ $l$-conformal Galilei superalgebra}
\vskip 0.5 cm
There exist two types of $\,\mathcal{N}=2$ supersymmetric extensions of the $l$-conformal Galilei algebra \cite{Masterov_1}, \cite{Aizawa_2} which are called real and chiral. Apart form the generators $K_{-1}$, $K_{1}$, $G_{-\frac{1}{2}}$, and $\bar{G}_{-\frac{1}{2}}$ mentioned above and the standard space rotations $M_{ij}$, both superalgebras involve the generator of dilatations $K_{0}$, the generators of superconformal transformations $G_{\frac{1}{2}}$, $\bar{G}_{\frac{1}{2}}$, the generator of $U(1)$ $R$-symmetry transformations $J$, the set of bosonic vector generators $C_i^{(n)}$ with $n=0,1,..,2l$, the set of fermionic vector generators $L_i^{(n)}$, $\bar{L}_i^{(n)}$ with $n=0,1,..,2l-1$, and the additional bosonic vector generators $P_i^{(n)}$ with $n=0,1,..,2l-2\gamma$, where
\bea\label{gamma}
\gamma=\left\{
\begin{aligned}
0&\qquad\mbox{for chiral superalgebra;}\\
1&\qquad\mbox{for real superalgebra.}\\
\end{aligned}
\right.
\eea
Taking into account the notation (\ref{gamma}), (anti)commutation relations of both the superalgebras can be written in the form
\bea\label{N=2GA}
&&
\{G_{r},\bar{G}_{s}\}=2iK_{r+s}+(-1)^{r+1/2}\d_{r+s,0}\,J,\qquad\;\, [J,C_i^{(n)}]=\;2l(1-\gamma)\, P_i^{(n)},
\nonumber
\\[5pt]
&&
[G_{r}, C_i^{(n)}]=(n-2l(r+1/2))L_{i}^{(n+r-1/2)},\quad\;\;\, [J,L_i^{(n)}]=\;\;\,i (1+2l(1-\gamma))\, L_i^{(n)},
\nonumber
\\[5pt]
&&
[\bar{G}_{r}, C_i^{(n)}]=(n-2l(r+1/2))\bar{L}_{i}^{(n+r-1/2)},\quad\;\;\, [J,\bar{L}_i^{(n)}]=-i(1+2l(1-\gamma))\,\bar{L}_i^{(n)},
\nonumber
\\[5pt]
&&
 [K_{m},C^{(n)}_i]=(n-l(m+1))C_i^{(n+m)},\qquad\quad\;\;\,[K_{p},K_{m}]=(m-p)K_{p+m},
\nonumber
\\[5pt]
&&
[K_{m},P^{(n)}_i]=(n-(l-\gamma)(m+1))P_i^{(n+m)},\quad\; [J,P_i^{(n)}]=-2l(1-\gamma)\,C_i^{(n)},
\nonumber
\\[5pt]
&&
[K_m,L_{i}^{(n)}]=(n-(l-1/2)(m+1))L_{i}^{(n+m)},\;\; [K_m,G_{r}]=(r-m/2)G_{m+r},
\nonumber
\\[5pt]
&&
[K_m,\bar{L}_{i}^{(n)}]=(n-(l-1/2)(m+1))\bar{L}_{i}^{(n+m)},\;\; [K_m,\bar{G}_{r}]=(r-m/2)\bar{G}_{m+r},
\nonumber
\eea
\bea
&&
[G_{r},P_i^{(n)}]=\;\;\, i\Bigl(1+(1-\gamma)[n-2l(r+1/2)-1]\Bigr)L_{i}^{(n+r+\gamma-1/2)},\;[J,G_{r}]=\;\;\,i\, G_{r},
\nonumber
\\[5pt]
&&
[\bar{G}_{r},P_i^{(n)}]=-i\Bigl(1+(1-\gamma)[n-2l(r+1/2)-1]\Bigr)\bar{L}_{i}^{(n+r+\gamma-1/2)},\;[J,\bar{G}_{r}]=-i\, \bar{G}_{r},
\nonumber
\\[5pt]
&&
\{G_{r},\bar{L}_{i}^{(n)}\}=i C_i^{(n+r+1/2)}-\Bigl(1+\gamma[n-(2l-1)(r+1/2)-1]\Bigr) P^{(n-\gamma+r+1/2)}_{i},
\nonumber
\\[5pt]
&&
\{\bar{G}_{r},L_{i}^{(n)}\}=i C_i^{(n+r+1/2)}+\Bigl(1+\gamma[n-(2l-1)(r+1/2)-1]\Bigr) P^{(n-\gamma+r+1/2)}_{i},
\nonumber
\\[5pt]
&&
[M_{ij},M_{ks}]=\delta_{ik} M_{js}+\delta_{js} M_{ik}-\delta_{is} M_{jk}-\delta_{jk} M_{is},
\nonumber
\\[5pt]
&&
[M_{ij},A^{(n)}_k]=\delta_{ik} A^{(n)}_j-\delta_{jk} A^{(n)}_i,\qquad A_i^{(n)}=C_i^{(n)},\;L_i^{(n)},\;\bar{L}_i^{(n)},\; P_i^{(n)}.
\eea

In contrast to the relation between the $l$-conformal Galilei algebra and its NH counterpart discussed around Eq. (\ref{change_2}) in the Introduction, for the supersymmetric extension one can choose two alternative bases in the $\,\mathcal{N}=2$ $l$-conformal Galilei superalgebra (\ref{N=2GA}) which are linked to different dynamical realizations in NH spacetime with negative cosmological constant. The first one involves the following change of the basis (see Ref. \cite{Masterov_3}):
\bea\label{red_1}
K_{-1}\rightarrow K_{-1}+\frac{1}{R^2}K_{1},\quad G_{-\frac{1}{2}}\rightarrow G_{-\frac{1}{2}}+\frac{i}{R}G_{\frac{1}{2}},\quad \bar{G}_{-\frac{1}{2}}\rightarrow\bar{G}_{-\frac{1}{2}}-\frac{i}{R}\bar{G}_{\frac{1}{2}}.
\eea
In particular, in Ref. \cite{Masterov_3} a generalization of the Pais-Uhlenbeck oscillator whose symmetries form an $\,\mathcal{N}=2$ $l$-conformal NH superalgebra in the this basis was constructed. The redefinition (\ref{red_1}) implies the following (anti)commutation relations between $K_{-1}$, $G_{-\frac{1}{2}}$, and $\bar{G}_{-\frac{1}{2}}$
\bea\label{nonstandard}
\begin{aligned}
&
\{G_{-\frac{1}{2}},\bar{G}_{-\frac{1}{2}}\}=2i\left(K_{-1}-\frac{1}{R}J\right),\quad[K_{-1},G_{-\frac{1}{2}}]=\frac{i}{R}G_{-\frac{1}{2}},
\quad [K_{-1},\bar{G}_{-\frac{1}{2}}]=-\frac{i}{R}\bar{G}_{-\frac{1}{2}},
\\[5pt]
&
\qquad\qquad\qquad\qquad\{G_{-\frac{1}{2}},G_{-\frac{1}{2}}\}=0,\qquad\qquad\qquad \{\bar{G}_{-\frac{1}{2}},\bar{G}_{-\frac{1}{2}}\}=0.
\end{aligned}
\eea
According to Ref. \cite{Toppan} these structure relations correspond to the so-called weak supersymmetry \cite{Smilga_3}.

In Ref. \cite{Masterov_1} another basis for the NH partner of the superalgebra (\ref{N=2GA}) was introduced. It relies upon redefinition of the following generators:
\bea\label{red_2}
K_{-1}\rightarrow K_{-1}+\frac{1}{R^2}K_{1}-\frac{1}{R}J,\quad G_{-\frac{1}{2}}\rightarrow G_{-\frac{1}{2}}+\frac{i}{R}G_{\frac{1}{2}},\quad \bar{G}_{-\frac{1}{2}}\rightarrow\bar{G}_{-\frac{1}{2}}-\frac{i}{R}\bar{G}_{\frac{1}{2}},
\eea
according to which the (anti)commutation relations between the generators $K_{-1}$, $G_{-\frac{1}{2}}$, and $\bar{G}_{-\frac{1}{2}}$ take the conventional form
\bea\label{standard}
\begin{aligned}
&
\{G_{-\frac{1}{2}},\bar{G}_{-\frac{1}{2}}\}=2iK_{-1}, \qquad [K_{-1},G_{-\frac{1}{2}}]=0,\qquad [K_{-1},\bar{G}_{-\frac{1}{2}}]=0,
\\[5pt]
&
\qquad\qquad\quad\;\{G_{-\frac{1}{2}},G_{-\frac{1}{2}}\}=0,\qquad \{\bar{G}_{-\frac{1}{2}},\bar{G}_{-\frac{1}{2}}\}=0.
\end{aligned}
\eea
It should be mentioned that the redefinitions (\ref{red_1}) and (\ref{red_2}) produce different expressions for the generator $K_{-1}$. This generator, however, corresponds to the Hamiltonian of a system if dynamical realizations are concerned. So, two variants of the basis (\ref{red_1}) and (\ref{red_2}) result in nonequivalent dynamical systems. But the NH basis of the superalgebra (\ref{N=2GA}) corresponding to (\ref{red_2}) is more appropriate for quantum mechanical applications in higher-derivative mechanics (see the discussion in Introduction). In subsequent sections, we study in detail the NH version of the superalgebra (\ref{N=2GA}) originating from the redefinition (\ref{red_2}) and construct its dynamical realizations.

\vskip 0.5 cm
\noindent
{\bf 3. Representation by the linear differential operators}
\vskip 0.5 cm

In a recent work \cite{Masterov_1}, a realization of the real $\,\mathcal{N}=2$ $l$-conformal NH superalgebra in terms of linear differential operators acting in NH spacetime has been constructed. It relied upon the Niederer-type transformation
\bea\label{nied_1}
t=R\tan{\frac{\mathit{\tau}}{R}},\qquad x_i=\chi_i/\cos^{2l}{\frac{\tau}{R}},\qquad \theta=\vartheta/\cos{\frac{\tau}{R}},\qquad \bar{\theta}=\bar{\vartheta}/\cos{\frac{\tau}{R}},
\eea
applied to the representation of real $\,\mathcal{N}=2$ $l$-conformal Galilei superalgebra (\ref{N=2GA}) \cite{Masterov_1}
\bea\label{GA}
&&
K_{n}=t^{n+1}\frac{\partial}{\partial t}+l(n+1)t^{n}x_i\frac{\partial}{\partial x_i}+\frac{1}{2}(n+1)t^{n}\theta\frac{\partial}{\partial \theta} +\frac{1}{2}(n+1)t^{n}\bar{\theta}\frac{\partial}{\partial \bar{\theta}},
\nonumber
\\[7pt]
&&
G_{-\frac{1}{2}}=i\frac{\partial}{\partial\bar{\theta}}+\theta\frac{\partial}{\partial t},\qquad G_{\frac{1}{2}}=i t\frac{\partial}{\partial\bar{\theta}}+\theta t\frac{\partial}{\partial t}+2l\theta x_i\frac{\partial}{\partial x_i}+\theta\bar{\theta}\frac{\partial}{\partial\bar{\theta}},
\nonumber
\\[7pt]
&&
\bar{G}_{-\frac{1}{2}}=i\frac{\partial}{\partial\theta}+\bar{\theta}\frac{\partial}{\partial t},\qquad \bar{G}_{\frac{1}{2}}=i t\frac{\partial}{\partial\theta}+\bar{\theta} t\frac{\partial}{\partial t}+2l\bar{\theta} x_i\frac{\partial}{\partial x_i}+\bar{\theta}\theta\frac{\partial}{\partial\theta},
\\[7pt]
&&
J=i\theta\frac{\partial}{\partial\theta}-i\bar{\theta}\frac{\partial}{\partial\bar{\theta}},\qquad\;\, M_{ij}=x_i\frac{\partial}{\partial x_j}-x_j\frac{\partial}{\partial x_i},
\nonumber
\\[7pt]
&&
C_i^{(n)}=t^{n}\frac{\partial}{\partial x_i},\quad L_i^{(n)}=\theta t^{n}\frac{\partial}{\partial x_i},\quad \bar{L}_i^{(n)}=\bar{\theta}t^{n}\frac{\partial}{\partial x_i},\quad P_i^{(n)}=\bar{\theta}\theta t^{n}\frac{\partial}{\partial x_i},
\nonumber
\eea
where $t,x_i$/$\tau,\chi_i$ designate even coordinates and $\theta,\bar\theta$/$\vartheta,\bar{\vartheta}$ denote odd coordinates parameterizing flat/Newton-Hooke superspace, respectively.
For the transformed operators $K_{-1}$, $K_{1}$, and $J$ one finds \cite{Masterov_1}
\bea\label{1}
\begin{aligned}
&
K_{-1}=\cos^{2}{\frac{\tau}{R}}\frac{\partial}{\partial\tau}-\frac{l}{R}\sin{\frac{2\tau}{R}}\chi_i\frac{\partial}{\partial\chi_i} -\frac{1}{2R}\sin{\frac{2\tau}{R}}\vartheta\frac{\partial}{\partial\vartheta}-\frac{1}{2R}\sin{\frac{2\tau}{R}}\bar{\vartheta} \frac{\partial}{\partial\bar{\vartheta}},
\\[5pt]
&
K_{1}=R^{2}\sin^{2}{\frac{\tau}{R}}\frac{\partial}{\partial\tau}+l R\sin{\frac{2\tau}{R}}\chi_i\frac{\partial}{\partial\chi_i}
+\frac{R}{2}\sin{\frac{2\tau}{R}}\vartheta\frac{\partial}{\partial\vartheta}+\frac{R}{2}\sin{\frac{2\tau}{R}}\bar{\vartheta} \frac{\partial}{\partial\bar{\vartheta}},\,J=i\theta\frac{\partial}{\partial\theta}-i\bar{\theta}\frac{\partial}{\partial\bar{\theta}}.
\end{aligned}
\eea
Then the change of the basis (\ref{red_2}) produces a representation of the $\,\mathcal{N}=2$ $l$-conformal NH superalgebra, a specific feature of which is that the operator $K_{-1}$ has the following form \cite{Masterov_1}:
\bea
K_{-1}=\frac{\partial}{\partial\tau}-\frac{1}{R}\left(i\vartheta\frac{\partial}{\partial\vartheta}-
i\bar{\vartheta}\frac{\partial}{\partial\bar{\vartheta}}\right).
\nonumber
\eea
In this case $K_{-1}$ is a linear combination of the generator of time translations and the generator of $U(1)$ $R$-symmetry transformation\footnote{Note that in the basis (\ref{red_1}) one obtains from Eq. (\ref{1}) the generator $K_{-1}$ in the form of the pure time translation. By this reason Niederer's transformations (\ref{nied_1}) can be naturally viewed as corresponding to the representation of the $\,\mathcal{N}=2$ $l$-conformal Galilei superalgebra in the basis (\ref{red_1}).}.

In order to preserve the (anti)commutation relations and to simultaneously transform the generator $K_{-1}$ to the form of the generator of pure time translations, let us deform the coordinate transformations (\ref{nied_1}) for the odd coordinates in the following way:
\bea
\theta=f(\tau)\vartheta/\cos{\frac{\tau}{R}},\qquad \bar{\theta}=g(\tau)\bar{\vartheta}/\cos{\frac{\tau}{R}},
\nonumber
\eea
Applying this Ansatz together with the transformations for the even coordinates in (\ref{nied_1}) to the generators (\ref{GA}), one gets
\bea
K_{-1}=\frac{\partial}{\partial\tau}-\left(\frac{1}{f(\tau)}\frac{d f(\tau)}{d\tau}+\frac{i}{R}\right)\vartheta\frac{\partial}{\partial\vartheta} -\left(\frac{1}{g(\tau)}\frac{d g(\tau)}{d\tau}-\frac{i}{R}\right)\bar{\vartheta}\frac{\partial}{\partial\bar{\vartheta}},
\nonumber
\eea
provided the redefinition (\ref{red_2}) has been done. In order to achieve our goal, it is sufficient to set $f(\tau)=e^{-\frac{i\tau}{R}}$, $g(\tau)=e^{\frac{i\tau}{R}}$. Other generators in the superalgebra which are obtained by applying the coordinate transformations to (\ref{GA}) read
\bea\label{NH}
\begin{aligned}
&
K_{0}=\frac{R}{2}\sin{\frac{2\tau}{R}}\frac{\partial}{\partial\tau}+l\cos{\frac{2\tau}{R}}\chi_i\frac{\partial}{\partial\chi_i} +\frac{1}{2}e^{\frac{2i\tau}{R}}\vartheta\frac{\partial}{\partial\vartheta}+\frac{1}{2}e^{-\frac{2i\tau}{R}}\bar{\vartheta}\frac{\partial}{\partial\bar{\vartheta}},
\\[7pt]
&
K_{1}=R^{2}\sin^{2}{\frac{\tau}{R}}\frac{\partial}{\partial\tau}+l R\sin{\frac{2\tau}{R}}\chi_i\frac{\partial}{\partial\chi_i}
+R\sin{\frac{\tau}{R}}e^{\frac{i\tau}{R}}\vartheta\frac{\partial}{\partial\vartheta}+R\sin{\frac{\tau}{R}}e^{-\frac{i\tau}{R}}\bar{\vartheta} \frac{\partial}{\partial\bar{\vartheta}},
\\[7pt]
&
G_{-\frac{1}{2}}=i\frac{\partial}{\partial\bar{\vartheta}}+\vartheta\frac{\partial}{\partial\tau}+ \frac{2il}{R}\vartheta\chi_i\frac{\partial}{\partial\chi_i},\quad \bar{G}_{-\frac{1}{2}}=i\frac{\partial}{\partial\vartheta}+\bar{\vartheta}\frac{\partial}{\partial\tau}- \frac{2il}{R}\bar{\vartheta}\chi_i\frac{\partial}{\partial\chi_i},
\\[7pt]
&
G_{\frac{1}{2}}=iR\sin{\frac{\tau}{R}}e^{-\frac{i\tau}{R}}\frac{\partial}{\partial\bar{\vartheta}}+R\sin{\frac{\tau}{R}}e^{-\frac{i\tau}{R}}\vartheta\frac{\partial}{\partial\tau}+
2l\cos{\frac{\tau}{R}}e^{-\frac{i\tau}{R}}\vartheta\chi_i\frac{\partial}{\partial\chi_i}+
e^{-\frac{2i\tau}{R}}\vartheta\bar{\vartheta}\frac{\partial}{\partial\bar{\vartheta}},
\\[7pt]
&
\bar{G}_{\frac{1}{2}}=iR\sin{\frac{\tau}{R}}e^{\frac{i\tau}{R}}\frac{\partial}{\partial\vartheta}+R\sin{\frac{\tau}{R}}e^{\frac{i\tau}{R}}\bar{\vartheta}\frac{\partial}{\partial\tau}+
2l\cos{\frac{\tau}{R}}e^{\frac{i\tau}{R}}\bar{\vartheta}\chi_i\frac{\partial}{\partial\chi_i}+
e^{\frac{2i\tau}{R}}\bar{\vartheta}\vartheta\frac{\partial}{\partial\vartheta},
\\[7pt]
&
C_i^{(n)}=R^{n}\sin^{n}{\frac{\tau}{R}}\cos^{2l-n}{\frac{\tau}{R}}\frac{\partial}{\partial \chi_i},\qquad\quad L_i^{(n)}=e^{-\frac{i\tau}{R}}R^{n}\sin^{n}{\frac{\tau}{R}}\cos^{2l-n-1}{\frac{\tau}{R}}\vartheta\frac{\partial}{\partial\chi_i},
\\[7pt]
&
P_i^{(n)}=R^{n}\sin^{n}{\frac{\tau}{R}}\cos^{2l-n-2}{\frac{\tau}{R}}\bar{\vartheta}\vartheta\frac{\partial}{\partial \chi_i}, \quad \bar{L}_i^{(n)}=e^{\frac{i\tau}{R}}R^{n}\sin^{n}{\frac{\tau}{R}}\cos^{2l-n-1}{\frac{\tau}{R}}\bar{\vartheta}\frac{\partial}{\partial\chi_i},
\\[7pt]
&
J=i\vartheta\frac{\partial}{\partial\vartheta}-i\bar{\vartheta}\frac{\partial}{\partial\bar{\vartheta}},\qquad M_{ij}=\chi_i\frac{\partial}{\partial\chi_{j}}-\chi_j\frac{\partial}{\partial\chi_{i}},
\end{aligned}
\eea
These generators obey the structure relations of real $\,\mathcal{N}=2$ $l$-conformal NH superalgebra. Note that in the flat space limit $R\rightarrow\infty$ the operators (\ref{NH}) reduce to those of $\,\mathcal{N}=2$ $l$-conformal Galilei superalgebra (\ref{GA}), as they should.

\vskip 0.5cm
\noindent
{\bf 4. Novel dynamical realizations}
\vskip 0.5cm
The simplest dynamical realization of $\,\mathcal{N}=2$ $l$-conformal Galilei superalgebra (\ref{N=2GA}) has been constructed in Ref. \cite{Masterov_3}. It describes the model of a superparticle which is governed by the action functional\footnote{The bosonic limit of (\ref{action}) has been recently studied in Refs. \cite{Horvathy}, \cite{Gomis}, \cite{Gonera_5}-\cite{Gonera_2}.}
\bea\label{action}
S=\frac{1}{2}\int\,dt\,\lambda_{ij}\,\left(x_i\frac{d^{2l+1}x_j}{dt^{2l+1}}- i\psi_i\frac{d^{2l}\bar{\psi}_j}{dt^{2l}}- i\bar{\psi}_i\frac{d^{2l}\psi_j}{dt^{2l}}+(-1)^{\gamma}\,z_i\frac{d^{2l-2\gamma+1}z_j}{dt^{2l-2\gamma+1}}\right),
\eea
where
\bea
\lambda_{ij}=\left\{
\begin{aligned}
\d_{ij},&\qquad i,j=1,2,..,d,&\qquad\mbox{for half-integer $l$};\\
\epsilon_{ij},&\qquad i,j=1,2,&\qquad\mbox{for integer $l$},\\
\end{aligned}
\right.
\eea
with $\epsilon_{12}=1$. $\gamma$ is defined in (\ref{gamma}).
The configuration space of the superparticle (\ref{action}) involves  the bosonic coordinates $x_i$, the fermionic coordinates $\psi_i$, $\bar{\psi}_i$, and the extra bosonic coordinates\footnote{For the dynamical realization corresponding to real $\,\mathcal{N}=2$ $l=1/2$-conformal Galilei superalgebra the coordinates $z_i$ and the generators $P_i^{(n)}$ are absent.} $z_i$.

The NH counterpart of (\ref{action}) has been also constructed in Ref. \cite{Masterov_3} by applying the following Niederer's transformation:
\bea\label{nied_2}
&&
t=R\tan(\tau/R),\qquad x_i(t)=\chi_i(\tau)/\cos^{2l}(\tau/R),\qquad z_i(t)=\pi_i(\tau)/\cos^{2l-2\gamma}(\tau/R),
\\[7pt]
&&\label{nied_8}
\qquad\qquad\psi_i(t)=\vartheta_i(\tau)/\cos^{2l-1}(\tau/R),\qquad \bar{\psi}_i(t)=\bar{\vartheta}_i(\tau)/\cos^{2l-1}(\tau/R).
\eea
However, as was mentioned above, the generators of the symmetry transformations corresponding to $K_{-1}$, $G_{-\frac{1}{2}}$ and $\bar{G}_{-\frac{1}{2}}$ obey the structure relations (\ref{nonstandard}). In Ref. \cite{Masterov_3} the generator of time translations has been obtained by applying the redefinition (\ref{red_1}) after implementing (\ref{nied_2}) (for more details, see Ref. \cite{Masterov_3}). If one is able to find
a coordinate transformation which results in the generator of time translations $K_{-1}$ after the redefinition (\ref{red_1}), one automatically obtains a dynamical realization of the superalgebra with the (anti)commutation relations (\ref{standard}). Let us obtain this coordinate transformation along the lines of the previous section.

For the case of the real superalgebra, the action functional (\ref{action}) is invariant under the symmetry transformations corresponding to $K_{-1}$, $K_{1}$, and $J$ in the form \cite{Masterov_3}
\bea\label{gen}
\begin{aligned}
&
K_{-1}=\frac{\partial}{\partial t},\qquad J=i\psi_i\frac{\partial}{\partial\psi_i}-i\bar{\psi}_i\frac{\partial}{\partial\bar{\psi}_i},
\\[5pt]
&
K_{1}=t^{2}\frac{\partial}{\partial t}+2ltx_i\frac{\partial}{\partial x_i}+(2l-1)t\psi_i\frac{\partial}{\partial \psi_i} +(2l-1)t\bar{\psi}_i\frac{\partial}{\partial\bar{\psi}_i}+(2l-2)t z_i\frac{\partial}{\partial z_i}.
\end{aligned}
\eea
Let us deform the coordinate transformations (\ref{nied_8}) for the fermionic degrees of freedom as follows:
\bea\label{deform}
\psi_i(t)=f(\tau)\vartheta_i(\tau)/\cos^{2l-1}(\tau/R),\qquad\bar{\psi}_i(t)=g(\tau)\bar{\vartheta}_i(\tau)/\cos^{2l-1}(\tau/R)
\eea
This deformation together with (\ref{nied_2}) yields
\bea
K_{-1}=\frac{\partial}{\partial \tau}-\left(\frac{1}{f}\frac{d f}{d\tau}+\frac{i}{R}\right)\vartheta_i\frac{\partial}{\partial\vartheta_i} -\left(\frac{1}{g}\frac{d g}{dt}-\frac{i}{R}\right)\bar{\vartheta}_i\frac{\partial}{\partial\bar{\vartheta}_i}
\nonumber
\eea
where is is assumed that the linear change of the basis (\ref{red_2}) has been implemented. So, Niederer's transformation
\bea\label{nied_3}
\begin{aligned}
&
t=R\tan{\frac{\tau}{R}},\qquad x_i(t)=\chi_i(\tau)/\cos^{2l}{\frac{\tau}{R}},\qquad \psi_i(t)=e^{-\frac{i\tau}{R}}\,\vartheta_i(\tau)/\cos^{2l-1}{\frac{\tau}{R}},
\\[2pt]
&
\qquad\qquad \bar{\psi}_i(t)=e^{\frac{i\tau}{R}}\,\bar{\vartheta}_i(\tau)/\cos^{2l-1}{\frac{\tau}{R}},
\qquad z_i(t)=\pi_i(\tau)/\cos^{2l-2}{\frac{\tau}{R}},
\end{aligned}
\eea
produces a dynamical realization of $\,\mathcal{N}=2$ $l$-conformal NH algebra which is the counterpart of (\ref{action}). For half-integer $l$, the application of (\ref{nied_3}) to (\ref{action}) gives
\bea\label{PU3}
\begin{aligned}
&
S=\frac{1}{2}\int d\tau\left(\chi_i\prod_{k=1}^{l+\frac{1}{2}}\left(\frac{d^2}{d\tau^2}+\frac{(2k-1)^2}{R^2}\right)\chi_i -i\vartheta_i\left(\frac{d}{d\tau}+\frac{2il}{R}\right)\prod_{k=1}^{l-\frac{1}{2}}\left(\frac{d^2}{d\tau^2}+\frac{(2k-1)^2}{R^2}\right)
\bar{\vartheta}_i\right.
\\[7pt]
&
\qquad\;\left.-i\bar{\vartheta}_i\left(\frac{d}{d\tau}-\frac{2il}{R}\right)\prod_{k=1}^{l-\frac{1}{2}}\left(\frac{d^2}{d\tau^2}+\frac{(2k-1)^2}{R^2}\right)
\vartheta_i-\,\pi_i\prod_{k=1}^{l-\frac{1}{2}}\left(\frac{d^2}{d\tau^2}+\frac{(2k-1)^2}{R^2}\right)\pi_i\right).
\end{aligned}
\eea
For integer $l$, one finds
\bea\label{PU4}
\begin{aligned}
&
S=\frac{1}{2}\int dt\, \epsilon_{ij}\left(\chi_i\prod_{k=1}^{l}\left(\frac{d^2}{d\tau^2}+\frac{(2k)^2}{R^2}\right)\dot{\chi}_j-i\vartheta_i\left(\frac{d}{d\tau}+\frac{2il}{R}\right) \prod_{k=1}^{l-1}\left(\frac{d^2}{dt^2}+\frac{(2k)^2}{R^2}\right)
\dot{\bar{\vartheta}}_j\right.
\\[7pt]
&
\qquad\,\left.-i\bar{\vartheta}_i\left(\frac{d}{d\tau}-\frac{2il}{R}\right)\prod_{k=1}^{l-1}\left(\frac{d^2}{dt^2}+\frac{(2k)^2}{R^2}\right)
\dot{\vartheta}_j-
\,\pi_i\prod_{k=1}^{l-1}\left(\frac{d^2}{d\tau^2}+\frac{(2k)^2}{R^2}\right)\dot{\pi}_j\right),
\end{aligned}
\eea
where the formal symbol $\prod\limits_{k=1}^{0}f(k)$  is assumed to be equal to the identity operator. These actions describe $\,\mathcal{N}=2$ supersymmetric extensions of the Pais-Uhlenbeck oscillator \cite{Pais} for the particular choice of its frequencies. Recently, the bosonic limit of (\ref{PU3}), (\ref{PU4}) has been extensively studied in Refs. \cite{PU}, \cite{Andrzejewski}, \cite{Galajinsky_1}, \cite{Andrzejewski_1}. An $\mathcal{N}=1$ superconformal extension of the Pais-Uhlenbeck oscillator has been investigated in Ref. \cite{Masterov_2}.

The difference between the systems (\ref{PU3}), (\ref{PU4}) and their analogues obtained in a recent work \cite{Masterov_3} concerns the dynamics of the odd degrees of freedom which obey to the following equations of motion:
\bea
\begin{aligned}
&
\qquad\qquad\qquad\qquad\qquad\quad\mbox{{\it the systems} (\ref{PU3}), (\ref{PU4})}\qquad\qquad\quad\quad\mbox{{\it the systems obtained in Ref.} \cite{Masterov_3}}
\\[5pt]
&
\mbox{half-integer $l$:}\quad
\begin{aligned}
\left(\frac{d}{d\tau}-\frac{2il}{R}\right)\prod_{k=1}^{l-\frac{1}{2}}\left(\frac{d^2}{d\tau^2}+\frac{(2k-1)^2}{R^2}\right)
\vartheta_i=0,\\
\left(\frac{d}{d\tau}+\frac{2il}{R}\right)\prod_{k=1}^{l-\frac{1}{2}}\left(\frac{d^2}{d\tau^2}+\frac{(2k-1)^2}{R^2}\right)
\bar{\vartheta}_i=0;\\
\end{aligned}
\quad
\begin{aligned}
\prod_{k=1}^{l-\frac{1}{2}}\left(\frac{d^2}{d\tau^2}+\frac{(2k)^2}{R^2}\right)\dot{\vartheta}_i=0,\\
\prod_{k=1}^{l-\frac{1}{2}}\left(\frac{d^2}{d\tau^2}+\frac{(2k)^2}{R^2}\right)\dot{\bar{\vartheta}}_i=0;\\
\end{aligned}
\end{aligned}
\nonumber
\eea
\bea
\begin{aligned}
&
\mbox{integer $l$:}\qquad\quad
\begin{aligned}
\left(\frac{d}{d\tau}-\frac{2il}{R}\right)\prod_{k=1}^{l-1}\left(\frac{d^2}{d\tau^2}+\frac{(2k)^2}{R^2}\right)
\dot{\vartheta}_i=0,\\
\left(\frac{d}{d\tau}+\frac{2il}{R}\right)\prod_{k=1}^{l-1}\left(\frac{d^2}{d\tau^2}+\frac{(2k)^2}{R^2}\right)
\dot{\bar{\vartheta}}_i=0;\\
\end{aligned}
\qquad
\begin{aligned}
\prod_{k=1}^{l}\left(\frac{d^2}{d\tau^2}+\frac{(2k-1)^2}{R^2}\right)\vartheta_i=0,\\
\prod_{k=1}^{l}\left(\frac{d^2}{d\tau^2}+\frac{(2k-1)^2}{R^2}\right)\bar{\vartheta}_i=0.\\
\end{aligned}
\end{aligned}
\nonumber
\eea

The symmetry generators $K_{-1}$, $K_{1}$, and $J$ of the system (\ref{action}) for the chiral case have the form \cite{Masterov_3}
\bea\label{chiral}
\begin{aligned}
&
K_{-1}=\frac{\partial}{\partial t},\qquad J=2l z_i\frac{\partial}{\partial x_i}-2l x_i\frac{\partial}{\partial z_i}+ i(2l+1)\psi_i\frac{\partial}{\partial\psi_i}-i(2l+1)\bar{\psi}_i\frac{\partial}{\partial\bar{\psi}_i},
\\[5pt]
&
K_{1}=t^{2}\frac{\partial}{\partial t}+2ltx_i\frac{\partial}{\partial x_i}+(2l-1)t\psi_i\frac{\partial}{\partial \psi_i} +(2l-1)t\bar{\psi}_i\frac{\partial}{\partial\bar{\psi}_i}+2lt z_i\frac{\partial}{\partial z_i}.
\end{aligned}
\eea
Let us deform Niederer's transformation for the odd coordinates (\ref{nied_8}) as in (\ref{deform}), while for the transformation of the even coordinates we choose the Ansatz
\bea
&&
x_i(t)=\frac{q_{1}(\tau)\chi_i(\tau)+r_1(\tau)\pi_i(\tau)}{\cos^{2l}{\frac{\tau}{R}}},\qquad z_i(t)=\frac{q_{2}(\tau)\chi_i(\tau)+r_2(\tau)\pi_i(\tau)}{\cos^{2l}{\frac{\tau}{R}}},
\nonumber
\eea
with
\bea
\Delta=q_1(\tau)r_2(\tau)-q_2(\tau)r_1(\tau)\neq 0.
\nonumber
\eea
Implementation of these transformations to (\ref{chiral}) yields $K_{-1}$
\bea
&&
K_{-1}=\frac{\partial}{\partial \tau}-\left(\frac{1}{f}\frac{d f}{d\tau}+\frac{(2l+1)i}{R}\right)\vartheta_i\frac{\partial}{\partial\vartheta_i} -\left(\frac{1}{g}\frac{d g}{d\tau}-\frac{(2l+1)i}{R}\right)\bar{\vartheta}_i\frac{\partial}{\partial\bar{\vartheta}_i}+
\nonumber
\\[5pt]
&&
\qquad\qquad\;+\frac{1}{\Delta}\left(r_1\Big(\frac{d q_2}{d\tau}-\frac{2l}{R}q_1\Big)-r_2\Big(\frac{d q_1}{d\tau}+\frac{2l}{R}q_2\Big)\right)\chi_i\frac{\partial}{\partial \chi_i}-
\nonumber
\\[5pt]
&&
\qquad\qquad\;-\frac{1}{\Delta}\left(q_1\Big(\frac{d q_2}{d\tau}-\frac{2l}{R}q_1\Big)-
q_2\Big(\frac{d q_1}{d\tau}+\frac{2l}{R}q_2\Big)\right)\chi_i\frac{\partial}{\partial \pi_i}+
\nonumber
\\[5pt]
&&
\qquad\qquad\;+\frac{1}{\Delta}\left(r_1\Big(\frac{d r_2}{d\tau}-\frac{2l}{R}r_1\Big)-r_2\Big(\frac{d r_1}{d\tau}+\frac{2l}{R}r_2\Big)
\right)\pi_i\frac{\partial}{\partial \chi_i}-
\nonumber
\\[5pt]
&&
\qquad\qquad\;-\frac{1}{\Delta}\left(q_1\Big(\frac{d r_2}{d\tau}-\frac{2l}{R}r_1\Big)-
q_2\Big(\frac{d r_1}{d\tau}+\frac{2l}{R}r_2\Big)\right)\pi_i\frac{\partial}{\partial \pi_i},
\nonumber
\eea
provided the redefinition (\ref{red_2}) has been done.

Thus, one obtains the desired $K_{-1}$ if Niederer's transformation takes the form
\bea\label{nied_5}
&&
\quad\; t=R\tan{\frac{\tau}{R}},\qquad \psi_i(t)=\frac{e^{-\frac{i(2l+1)\tau}{R}}\vartheta_i(\tau)}{\cos^{2l-1}{\frac{\tau}{R}}},\qquad \bar{\psi}_i(t)=\frac{e^{\frac{i(2l+1)\tau}{R}}\bar{\vartheta}_i(\tau)}{\cos^{2l-1}{\frac{\tau}{R}}},
\\[7pt]
&&\label{nied_6}
x_i(t)=\frac{\chi_i(\tau)\cos{\frac{2l\tau}{R}}-\pi_i(\tau)\sin{\frac{2l\tau}{R}}}{\cos^{2l}{\frac{\tau}{R}}},\quad
\qquad z_i(t)=\frac{\pi_i(\tau)\cos{\frac{2l\tau}{R}}+\chi_i(\tau)\sin{\frac{2l\tau}{R}}}{\cos^{2l}{\frac{\tau}{R}}}.
\eea

The action functional, which is obtained by applying (\ref{nied_5}), (\ref{nied_6}) to (\ref{action}), can be written in a compact form
\bea\label{action_1}
\begin{aligned}
&
S=\frac{1}{2}\la_{ij}\int\,d\tau\left(\frac{1}{2}\xi_i\prod_{k=0}^{2l}\Big(\frac{d}{d\tau}-\frac{2ik}{R}\Big)\bar{\xi}_j+\frac{1}{2}
\bar{\xi}_i\prod_{k=0}^{2l}\Big(\frac{d}{d\tau}+\frac{2ik}{R}\Big)\xi_j-\right.
\\[2pt]
&
\qquad\qquad\qquad\;\left.-i\vartheta_i\prod_{k=1}^{2l}\Big(\frac{d}{d\tau}+\frac{2ik}{R}\Big)\bar{\vartheta}_j
-i\bar{\vartheta}_i\prod_{k=1}^{2l}\Big(\frac{d}{d\tau}-\frac{2ik}{R}\Big)\vartheta_j\right).
\end{aligned}
\eea
where we denoted
\bea\label{coord_1}
\xi_i=\chi_i+i\pi_i,\qquad \bar{\xi}_i=\chi_i-i\pi_i.
\eea
Note that,
in contrast to the dynamical realization of the chiral $\,\mathcal{N}=2$ $l$-conformal NH superalgebra with the non-standard (anti)commutation relations \cite{Masterov_3},
the system (\ref{action_1}) cannot be interpreted as a supersymmetric generalization of the Pais-Uhlenbeck oscillator. The dynamics of the variables $\xi_i$, $\bar{\xi}_i$, $\vartheta_i$, $\bar{\vartheta}_i$ is governed by the equations of motion
\bea
&&
\prod_{k=0}^{2l}\Big(\frac{d}{d\tau}+\frac{2ik}{R}\Big)\xi_i=0,\qquad\prod_{k=0}^{2l}\Big(\frac{d}{d\tau}-\frac{2ik}{R}\Big)\bar{\xi}_i=0,
\nonumber
\\[2pt]
&&
\prod_{k=1}^{2l}\Big(\frac{d}{d\tau}-\frac{2ik}{R}\Big)\vartheta_i=0,\qquad\prod_{k=1}^{2l}\Big(\frac{d}{d\tau}+\frac{2ik}{R}\Big)\bar{\vartheta}_i=0.
\nonumber
\eea

If one introduces analogues of (\ref{coord_1}) for the flat superspace
\bea\label{coord}
y_i=x_i+i z_i,\qquad \bar{y}_i=x_i-iz_i,
\eea
then the coordinate transformations (\ref{nied_6}) can be rewritten in the form
\bea\label{nied_7}
y_i(t)=\frac{\xi_i e^{\frac{2il\tau}{R}}}{\cos^{2l}{\frac{\tau}{R}}},\qquad
\bar{y}_i=\frac{\bar{\xi}_i e^{-\frac{2il\tau}{R}}}{\cos^{2l}{\frac{\tau}{R}}}.
\eea

The symmetry transformations of the model (\ref{action}) have been introduced in Ref. \cite{Masterov_3}. These transformations can be mapped into those of (\ref{PU3}), (\ref{PU4}), and (\ref{action_1}) by applying Niederer's transformations (\ref{nied_3}), (\ref{nied_5}), and (\ref{nied_7}), respectively. The explicit form of the symmetry transformations of the action functionals obtained in this section is presented in Appendix.

\vskip 0.5cm
\noindent
{\bf 5. Conclusion}
\vskip 0.5cm
To summarize, in this work we have constructed a realization of real $\,\mathcal{N}=2$ $l$-conformal NH superalgebra  for the case of negative cosmological constant in terms of linear differential operators. As compared to the realization in Ref. \cite{Masterov_1}, the operator $K_{-1}$ generates genuine time translations. Also we have obtained dynamical realizations of real and chiral variants of $\,\mathcal{N}=2$ $l$-conformal NH superalgebra for the case of negative cosmological constant. For the real case, an $\,\mathcal{N}=2$ supersymmetric generalization of the Pais-Uhlenbeck oscillator for the particular choice of its frequencies has been constructed. For the chiral case, we obtained a dynamical system whose bosonic sector admits an elegant formulation in terms of a pair of complex conjugate coordinates. Niederer-type transformation which links both the systems with the model of a free higher derivative superparticle \cite{Masterov_3} has been found.

The dynamical realizations which have been constructed in this work are classical and involve higher-derivative terms in the equations of motion for the $l>1$. It is interesting to study in detail whether supersymmetry can help to resolve the ghost problem intrinsic to the Pais-Uhlenbeck oscillator (\ref{PU3}), (\ref{PU4}) at the quantum level. This issue will be studied elsewhere \cite{Masterov_4}.

\vskip 0.5cm
\noindent
{\bf Acknowledgements}
\vskip 0.5cm
We thank A. Galajinsky for the comments on the manuscript. This work was supported
by the Dynasty Foundation, RFBR grant 14-02-31139-Mol, the MSE program "Nauka" under
the project 3.825.2014/K, and the TPU grant LRU.FTI.123.2014.

\vskip 0.5cm
\noindent
{\bf Appendix. Symmetry transformations of the invariant actions}
\vskip 0.5cm
\noindent

An application of Niederer's transformation (\ref{nied_3}) for the real case and its analogue (\ref{nied_5}), (\ref{nied_7}) for the chiral case to the symmetry transformations of the model (\ref{action}) results in the symmetry transformations of the action functionals (\ref{PU3}), (\ref{PU4}) and (\ref{action_1}), respectively. In particular, for the actions (\ref{PU3}) and (\ref{PU4}) one finds
\bea
\begin{aligned}
&
K_{-1}:\;\; \d\tau=a_{-1};\qquad\quad J:\quad \d\vartheta_i=i\vartheta_i\nu,\quad \d\bar{\vartheta}_i=-i\bar{\vartheta}_i\nu;
\\[12pt]
&
K_{0}:\quad \d \tau=\frac{R}{2}\sin{\frac{2\tau}{R}}a_{0},\quad\d \chi_i=l\cos{\frac{2\tau}{R}}\, \chi_i\,a_{0},\quad
\d\vartheta_i=((l-1/2)\cos{\frac{2\tau}{R}}+\frac{i}{2}\sin{\frac{2\tau}{R}})\,\vartheta_i\,a_{0},
\\[2pt]
&
\qquad\qquad\qquad \d\bar{\vartheta}_i=((l-1/2)\cos{\frac{2\tau}{R}}-\frac{i}{2}\sin{\frac{2\tau}{R}})\,\bar{\vartheta}_i\,a_{0},\quad
\d \pi_i=(l-1)\cos{\frac{2\tau}{R}}\,\pi_i\,a_{0};
\\[12pt]
&
K_{1}:\quad\d \tau=R^{2}\sin^{2}{\frac{\tau}{R}} a_{1},\quad\d \chi_i=lR\sin{\frac{2\tau}{R}}\,\chi_i, a_{1},\quad \d \pi_i=(l-1)R\sin{\frac{2\tau}{R}}\,\pi_i\, a_{1},
\\[2pt]
&
\qquad\qquad\qquad\qquad\quad\d\vartheta_i=\left(\Big(l-\frac{1}{2}\Big)R\sin{\frac{2\tau}{R}}+iR\sin^2{\frac{\tau}{R}}\right)\vartheta_i\, a_{1},
\\[2pt]
&
\qquad\qquad\qquad\qquad\quad\d\bar{\vartheta}_i=\left(\Big(l-\frac{1}{2}\Big)R\sin{\frac{2\tau}{R}}-iR\sin^2{\frac{\tau}{R}}\right)\bar{\vartheta}_i\, a_{1};
\\[12pt]
&
G_{-\frac{1}{2}}:\quad \d \chi_i=i\vartheta_i\alpha_{-\frac{1}{2}},\quad\d \pi_i=\left(-\dot{\vartheta}_i+\frac{2il}{R}\vartheta_i\right)\alpha_{-\frac{1}{2}},\quad \d\bar{\vartheta}_i=\left(\dot{\chi}_i-\frac{2il}{R}\chi_i+i\pi_i\right)\alpha_{-\frac{1}{2}};
\\[12pt]
&
\bar{G}_{-\frac{1}{2}}:\quad\d \chi_i=i\bar{\vartheta}_i\bar{\alpha}_{-\frac{1}{2}},\quad
\d \pi_i=\left(\dot{\bar{\vartheta}}_i+\frac{2il}{R}\bar{\vartheta}_i\right)\bar{\alpha}_{-\frac{1}{2}},\quad\;\;\,
\d\vartheta_i=\left(\dot{\chi}_i+\frac{2il}{R}\chi_i-i\pi_i\right)\bar{\alpha}_{-\frac{1}{2}};
\end{aligned}
\nonumber
\eea
\bea
\begin{aligned}
&
G_{\frac{1}{2}}:\quad\;\; \d \chi_i=iR\sin{\frac{\tau}{R}}e^{-\frac{i\tau}{R}}\,\vartheta_i\,\alpha_{\frac{1}{2}},\quad
\d\bar{\vartheta}_i=R\sin{\frac{\tau}{R}}e^{-\frac{i\tau}{R}}\left(\dot{\chi}_i-\frac{2l}{R}\cot{\frac{\tau}{R}}\chi_i
+i \pi_i\right)\alpha_{\frac{1}{2}},
\\[2pt]
&
\qquad\qquad\qquad\;\d \pi_i=e^{-\frac{i\tau}{R}}\left(-R\sin{\frac{\tau}{R}}\vartheta_i^{(1)}+\Big((2l-1)\cos{\frac{\tau}{R}}+i\sin{\frac{\tau}{R}}\Big)
\vartheta_i\right)\alpha_{\frac{1}{2}};
\\[12pt]
&
\bar{G}_{\frac{1}{2}}:\quad\;\;\d \chi_i=iR\sin{\frac{\tau}{R}}e^{\frac{i\tau}{R}}\,\bar{\vartheta}_i\,\bar{\alpha}_{\frac{1}{2}},\quad \d\vartheta_i=R\sin{\frac{\tau}{R}}e^{\frac{i\tau}{R}}\left(\dot{\chi}_i -\frac{2l}{R}\cot{\frac{\tau}{R}}\chi_i-i \pi_i\right)\bar{\alpha}_{\frac{1}{2}},
\\[2pt]
&
\qquad\qquad\qquad\;
\d \pi_i=e^{\frac{i\tau}{R}}\left(R\sin{\frac{\tau}{R}}\dot{\bar{\vartheta}}_i+\Big(-(2l-1)\cos{\frac{\tau}{R}}+i\sin{\frac{\tau}{R}}\Big)\bar{\vartheta}_i\right) \bar{\alpha}_{\frac{1}{2}};
\\[12pt]
&
C_i^{(n)}:\quad \d \chi_i=R^{n}\sin^{n}{\frac{\tau}{R}}\cos^{2l-n}{\frac{\tau}{R}}\, a_i^{(n)};\quad L_i^{(n)}:\quad\d\bar{\vartheta}_i=e^{-\frac{i\tau}{R}}R^{n}\sin^{n}{\frac{\tau}{R}}\cos^{2l-n-1}{\frac{\tau}{R}}\,\xi_i^{(n)};
\\[12pt]
&
\bar{L}_i^{(n)}:\quad\d\vartheta_i=e^{\frac{i\tau}{R}}R^{n}\sin^{n}{\frac{\tau}{R}}\cos^{2l-n-1}{\frac{\tau}{R}}\,\bar{\xi}_i^{(n)};\quad
P_i^{(n)}:\quad \d \pi_i=R^{n}\sin^{n}{\frac{\tau}{R}}\cos^{2l-n-2}{\frac{\tau}{R}}\, b_i^{(n)};
\\[12pt]
&
M_{ij}:\quad\,\d\chi_i=w_{ij}\chi_j,\quad \d\vartheta_i=w_{ij}\vartheta_j,\quad \d\bar{\vartheta}_i=w_{ij}\bar{\vartheta}_i,\quad \d\pi_i=w_{ij}\pi_j,\quad\mbox{with}\quad w_{ij}=-w_{ji}.
\end{aligned}
\nonumber
\eea
\vskip 0.1cm

For the action (\ref{action_1}) the symmetry transformations have the form
\bea
\begin{aligned}
&
K_{-1}:\;\,\d \tau=a_{-1};
\\[10pt]
&
K_{0}:\quad\, \d \tau=\frac{R}{2}\sin{\frac{2\tau}{R}}\,a_0,\qquad\qquad\;\, \d\xi_i=le^{-\frac{2i\tau}{R}}\xi_i\,a_0,\qquad\qquad\;\, \d\bar{\xi}_i=le^{\frac{2i\tau}{R}}\bar{\xi}_i\,a_0,
\\[2pt]
&
\qquad\qquad\qquad\qquad\,\d\vartheta_i=\left((l-1/2)\cos{\frac{2\tau}{R}}+i(l+1/2)\sin{\frac{2\tau}{R}}\right)\vartheta_i\,a_0,
\\[2pt]
&
\qquad\qquad\qquad\qquad\,\d\bar{\vartheta}_i=\left((l-1/2)\cos{\frac{2\tau}{R}}-i(l+1/2)\sin{\frac{2\tau}{R}}\right)\bar{\vartheta}_i\,a_0
\\[10pt]
&
K_{1}:\quad \d \tau=R^{2}\sin^{2}{\frac{\tau}{R}}\,a_1,\quad \d\xi_i=-ilR\left(1-e^{-\frac{2i\tau}{R}}\right)\xi_i\,a_1,\quad \d\bar{\xi}_i=ilR\left(1-e^{\frac{2i\tau}{R}}\right)\bar{\xi}_i\,a_1,
\\[2pt]
&
\qquad\qquad\qquad\quad\;\d\vartheta_i=(l-1/2)R\sin{\frac{2\tau}{R}}\,\vartheta_i\,a_1+i(2l+1)R\sin^{2}{\frac{\tau}{R}}\,\vartheta_i\,a_1,
\\[2pt]
&
\qquad\qquad\qquad\quad\;  \d\bar{\vartheta}_i=(l-1/2)R\sin{\frac{2\tau}{R}}\,\bar{\vartheta}_i\,a_1-i(2l+1)R\sin^{2}{\frac{\tau}{R}}\,\bar{\vartheta}_i\,a_1;
\\[10pt]
&
J:\quad\;\;\, \d\xi_i=-2il\xi_i,\qquad\d\bar{\xi}_i=2il\bar{\xi}_i,\qquad\d\vartheta_i=i(2l+1)\nu\vartheta_i,\qquad \d\bar{\vartheta}_i=-i(2l+1)\nu\bar{\vartheta}_i;
\\[10pt]
&
G_{-\frac{1}{2}}:\;\, \d\bar{\vartheta}_i=\dot{\xi}_i\alpha_{-\frac{1}{2}},\quad \d\bar{\xi}_i=2i\vartheta_i\alpha_{-\frac{1}{2}};\quad\;\,
\bar{G}_{-\frac{1}{2}}:\;\, \d\vartheta_i=\dot{\bar{\xi}}_i\bar{\alpha}_{-\frac{1}{2}},\quad \d\xi_i=2i\bar{\vartheta}_i\bar{\alpha}_{-\frac{1}{2}};
\\[10pt]
&
G_{\frac{1}{2}}:\quad \d\bar{\vartheta}_i=R\sin{\frac{\tau}{R}}e^{-\frac{i\tau}{R}}\dot{\xi}_i\alpha_{\frac{1}{2}}-
2l e^{-\frac{2i\tau}{R}}\xi_i\alpha_{\frac{1}{2}},\qquad \d\bar{\xi}_i=2i R\sin{\frac{\tau}{R}}e^{-\frac{i\tau}{R}}\vartheta_i \alpha_{\frac{1}{2}};
\\[10pt]
&
\bar{G}_{\frac{1}{2}}:\quad \d\vartheta_i=R\sin{\frac{\tau}{R}}e^{\frac{i\tau}{R}}\dot{\bar{\xi}}_i\bar{\alpha}_{\frac{1}{2}}-
2l e^{\frac{2i\tau}{R}}\bar{\xi}_i\bar{\alpha}_{\frac{1}{2}},\qquad\quad\, \d\xi_i=2i R\sin{\frac{\tau}{R}}e^{\frac{i\tau}{R}}\bar{\vartheta}_i \bar{\alpha}_{\frac{1}{2}};
\end{aligned}
\nonumber
\eea
\bea
\begin{aligned}
&
C_i^{(n)}:\;\, \d \xi_i=R^{n}\sin^{n}{\frac{\tau}{R}}\cos^{2l-n}{\frac{\tau}{R}}e^{-\frac{2il\tau}{R}}a_i^{(n)},\qquad\;\;\;\, \d \bar{\xi}_i=R^{n}\sin^{n}{\frac{\tau}{R}}\cos^{2l-n}{\frac{\tau}{R}}e^{\frac{2il\tau}{R}}a_i^{(n)};
\\[10pt]
&
L_i^{(n)}:\;\, \d\bar{\vartheta}_i=e^{-\frac{i(2l+1)\tau}{R}}R^{n}\sin^{n}{\frac{\tau}{R}}\cos^{2l-n-1}{\frac{\tau}{R}}\zeta_i^{(n)};
\\[10pt]
&
\bar{L}_i^{(n)}:\;\, \d\vartheta_i=e^{\frac{i(2l+1)\tau}{R}}R^{n}\sin^{n}{\frac{\tau}{R}}\cos^{2l-n-1}{\frac{\tau}{R}}\bar{\zeta}_i^{(n)};
\\[10pt]
&
P_i^{(n)}:\;\, \d \xi_i=iR^{n}\sin^{n}{\frac{\tau}{R}}\cos^{2l-n}{\frac{\tau}{R}}e^{-\frac{2il\tau}{R}}b_i^{(n)},\qquad\quad\,
\d \bar{\xi}_i=-iR^{n}\sin^{n}{\frac{\tau}{R}}\cos^{2l-n}{\frac{\tau}{R}}e^{\frac{2il\tau}{R}}b_i^{(n)}.
\\[10pt]
&
M_{ij}:\;\;\, \d \xi_i=w_{ij}\xi_j,\quad \d \bar{\xi}_i=w_{ij}\bar{\xi}_j,\quad \d\vartheta_i=w_{ij}\vartheta_j,\quad \d\bar{\vartheta}_i=w_{ij}\bar{\vartheta}_j, \quad\mbox{with}\quad w_{ij}=-w_{ji}.
\end{aligned}
\nonumber
\eea

\end{document}